\documentclass[article]{aa}

\usepackage{graphicx}
\usepackage{ae}
\usepackage{txfonts}
\usepackage{natbib}
\bibpunct{(}{)}{;}{a}{}{,}

\begin{document}

\title{The stellar mass spectrum from non-isothermal gravoturbulent fragmentation}


\author{Anne-Katharina\ Jappsen\inst{1} \and
  Ralf\ S.\ Klessen\inst{1} \and Richard\ B.\ Larson\inst{2} \and Yuexing\ Li
  \inst{3,4}  \and Mordecai-Mark\ Mac Low \inst{3,4}} \authorrunning{A.-K.\
  Jappsen et al.} \titlerunning{Mass spectrum from non-isothermal gravoturbulent fragmentation} \offprints{A.-K.\ 
  Jappsen, \email{akjappsen@aip.de}}

\institute{Astrophysikalisches Institut Potsdam, An der Sternwarte 16, D-14482
  Potsdam, Germany \and Department of Astronomy, Yale University, New Haven,
  CT 06520-8101, USA \and Department of Astrophysics, American Museum of Natural
  History, 79th Street at Central Park West, New York, NY 10024-5192, USA \and
  Department of Astronomy, Columbia University, New York, NY 10027, USA }

\date{Received <date> / Accepted <date>}
\abstract{
Identifying the processes that determine the initial mass
function of stars (IMF) is a fundamental problem in star formation
theory. One of the major uncertainties is the exact chemical state of
the star forming gas and its influence on the dynamical evolution.
Most simulations of star forming clusters use an isothermal equation
of state (EOS). However, this might be an oversimplification given the
complex interplay between heating and cooling processes in molecular
clouds.  Theoretical predictions and observations suggest that the
effective polytropic exponent $\gamma$ in the EOS varies with density.
 We address these issues and study the effect of a piecewise polytropic
EOS on the formation of stellar clusters in turbulent,
self-gravitating molecular clouds using three-dimensional, smoothed
particle hydrodynamics simulations. In these simulations stars form via a
process we call gravoturbulent fragmentation, i.e., gravitational
fragmentation of turbulent gas. To approximate the results of published
predictions of the thermal behavior of collapsing clouds, we increase the
polytropic exponent $\gamma$ from 0.7 to 1.1 at a critical density $n_{\rm c}$,
which we estimated to be $2.5\times10^5\,\mathrm{cm^{-3}}$. The change of thermodynamic state
at $n_{\rm c}$ selects a characteristic mass scale for fragmentation
$M_{\rm ch}$, which we relate to the peak of the observed IMF. A simple scaling argument based on the Jeans mass $M_\mathrm{J}$ at the
critical density $n_\mathrm{c}$ leads to $M_{\mathrm{ch}}\propto
n_{\mathrm{c}}^{-0.95}$.  We perform simulations with
$4.3\times10^4\,\mathrm{cm^{-3}} < n_{\mathrm{c}} <
4.3\times10^7\,\mathrm{cm^{-3}}$ to test this scaling argument. 
Our simulations qualitatively support this hypothesis, but we find a weaker
density dependence of $M_{\mathrm{ch}} \propto n_{\mathrm{c}}^{-0.5\pm0.1}$. 
We also investigate the influence of additional environmental
parameters on the IMF. We consider variations in the turbulent driving
scheme, and consistently find
$M_{\mathrm{J}}$ is decreasing with increasing $n_{\mathrm{c}}$.
Our investigation generally supports the idea that the distribution
of stellar masses depends mainly on the thermodynamic state of the
star-forming gas.  The thermodynamic state of interstellar gas is a
result of the balance between heating and cooling processes, which in
turn are determined by fundamental atomic and molecular physics and by
chemical abundances. Given the abundances, the
derivation of a characteristic stellar mass can thus be based on
universal quantities and constants. 

}

\maketitle
\section{Introduction}

One of the fundamental unsolved problems in astronomy is the origin of the
stellar mass spectrum, the so-called initial mass function
(IMF). Observations suggest that there is a characteristic mass for
stars in the solar vicinity. The IMF peaks at this characteristic mass which
is typically a few tenths of a solar mass. The IMF has a nearly power-law form
for larger masses and declines rapidly towards smaller masses \citep{SCA98a,
  KRO02, CHA03}.
 
Although the IMF has been derived from vastly different regions, from the solar
vicinity to dense clusters of newly formed stars, the basic features seem to
be strikingly universal to all determinations \citep{KRO01b}. Initial conditions in star
forming regions can vary considerably. If the IMF depends on the initial
conditions, there would thus be no reason for it to be universal.
Therefore a derivation of the characteristic stellar mass that is based on
fundamental atomic and molecular physics would be more consistent.

There have been analytical models \citep{JEA02, LAR69, PEN69, LOW76, SHU77, WHI85} and numerical investigations of the effects
of various physical processes on collapse and fragmentation. These
  processes include, for example, magnetic fields \citep{BAS95a, TOM96, GAL01},
  feedback from the stars themselves \citep{SIL95,NAK95,ADA96} and competitive
  coagulation or accretion \citep{SIL79, LEJ86, PRI95, MUR96, BON01a, BON01b,
    DUR01}. In another group of models, initial and environmental conditions,
  like the structural properties of molecular clouds determine the IMF
  \citep{ELM83, ELM97a, ELM97b, ELM99a, ELM00a, ELM00c, ELM02a}. \citet{LAR73a}
  and \citet{ZIN84, ZIN90} argued in a more statistical approach that the
  central-limit theorem naturally leads to a log-normal stellar mass
  spectrum. Moreover, there are models that connect turbulent motions in molecular clouds to the IMF
  \citep[e.g.][]{LAR81, FLE82, PAD95, PAD97, KLE98, KLE00b, KLE01c, PAD02}.
  The interplay between turbulent motion and self-gravity of the cloud leads to a
  process we call gravoturbulent fragmentation. The supersonic turbulence
  ubiquitously observed in molecular gas generates strong density fluctuations
  with gravity taking over in the densest and most massive regions. Once gas
  clumps become gravitationally unstable, collapse sets in. The central
  density increases until a protostellar objects form and grow in mass via
  accretion from the infalling envelope. 
  For more detailed reviews see \citet{LAR03} and \citet{MAC04} .  

However, current results are generally based on models that do not treat thermal physics
in detail. Typically, they use a simple equation of state (EOS) which is
isothermal with the polytropic exponent $\gamma=1$. The true nature of the
EOS remains a major theoretical problem in understanding the fragmentation
properties of molecular clouds. Some calculations invoke cooling during the
collapse \citep{MON91, TUR95, WHI95}. Others include radiation transport to
account for the heating that occurs once the cloud reaches densities of
$n(\mathrm{H}_2) \geq 10^{10}\,\mathrm{cm}^{-3}$ \citep{MYH92,BOS93}, or
simply assume an adiabatic equation of state once that density is exceeded
\citep{BON94, BAT95}. \citet{SPA00} showed that radiatively cooling
gas can be described by a piecewise polytropic EOS, in which the polytropic
exponent $\gamma$ changes with gas density $\rho$.    
Considering a polytropic EOS is still a rather crude approximation. In
practice the behaviour of $\gamma$ may be more complicated and
important effects like the temperature of the dust, line-trapping and feedback from
newly-formed stars should be taken into account \citep{SCA98b}. Nevertheless a
polytropic EOS gives an insight in the differences that a departure from isothermality evokes.

Recently \citet{LI03} conducted a systematic study of the 
effects of a varying polytropic exponent~$\gamma$ on gravoturbulent 
fragmentation. Their results showed that $\gamma$ determines 
how strongly self-gravitating gas fragments. They found that the degree of
fragmentation decreases with increasing polytropic exponent $\gamma$ in the
range $0.2 < \gamma < 1.4$ although the total amount of mass in collapsed
cores appears to remain roughly consistent through this range. These findings
suggest that the IMF might be quite sensitive to the thermal physics. 
Earlier, one-dimensional simulations by \citet{PAS98} already showed that the density probability distribution of supersonic
turbulent gas displays a dependence on the polytropic exponent $\gamma$.
However in 
both computations, $\gamma$ was left strictly constant in 
each case. In this study we extend previous work 
by using a piecewise polytropic equation of state changing 
$\gamma$ at some chosen density. We investigate if a change 
in $\gamma$ determines the characteristic mass of the gas 
clump spectrum and thus, possibly, the turn-over mass of the 
IMF. 
 
In Sect.~\ref{sec:therm-prop} we review what is currently known about the
thermal properties of interstellar gas. 
In Sect.~\ref{sec:ana-app} we approach the fragmentation problem analytically, while in
Sect.~\ref{sec:num} we introduce our computational method. In
Sect.~\ref{sec:res} we discuss gravoturbulent fragmentation in non-isothermal
gas. In Sect.~\ref{sec:char} we analyze the resulting mass
distribution. We further
investigate the influence of different turbulent driving fields and different
scale of driving in Sect.~\ref{sec:dep}.
Finally, in Sect.~\ref{sec:sum} we summarize. 

\section{Thermal Properties of Star-Forming Clouds}
\label{sec:therm-prop}
 Gravity in galactic molecular clouds is initially expected to be opposed
  mainly by a combination of supersonic turbulence and magnetic fields 
\citep{MAC04}. The velocity structure in the clouds is always observed to be
dominated by large-scale modes \citep{MAC00, OSS01, OSS02}. In order to
maintain turbulence for some global dynamical timescales and to compensate for
gravitational contraction of the cloud as a whole, kinetic energy input from
external sources seems to be required. Star formation then takes place in molecular cloud
regions which are characterized by local dissipation of turbulence and loss of
magnetic flux, eventually 
leaving thermal pressure as the main force resisting
gravity in the small dense prestellar cloud cores that actually build up the
stars \citep{KLE04, VAZ05}.  In agreement
with this expectation, observed prestellar cores typically show a rough
balance between gravity and thermal pressure \citep{BEN89, MYE91}.  Therefore the thermal
properties of the dense star-forming regions of molecular clouds must play
an important role in determining how these clouds collapse and fragment
into stars.

Early studies of the balance between heating and cooling processes in
collapsing clouds predicted temperatures of the order of $10\,$K to $20\,$K,
tending to be lower at the higher densities
\citep[e.g.,][]{HAY65,HAY66,LAR69,LAR73}. In their dynamical collapse calculations,
these and other authors approximated this somewhat varying temperature by
a simple constant value, usually taken to be $10\,$K.  Nearly all subsequent
studies of cloud collapse and fragmentation have used a similar isothermal
approximation.  However, this approximation is actually only a somewhat
crude one, valid only to a factor of 2, since the temperature is predicted
to vary by this much above and below the usually assumed constant value
of $10\,$K.  Given the strong sensitivity of the results of fragmentation
simulations like those of \citet{LI03} to the assumed
equation of state of the gas, temperature variations of this magnitude may
be important for quantitative predictions of stellar masses and the IMF.

As can be seen in Fig.~2 of \citet{LAR85}, observational and theoretical
studies of the thermal properties of collapsing clouds both indicate that
at densities below about $10^{-18}\,\mathrm{g\,cm}^{-3}$, roughly corresponding
to a number density of $n = 2.5\times 10^5\,\mathrm{cm}^{-3}$, the temperature generally
decreases with increasing density.  In this low-density regime, clouds are
externally heated by cosmic rays or photoelectric heating, and they are
cooled mainly by the collisional excitation of low-lying levels of C$^+$ ions and
O atoms; the strong dependence of the cooling rate on density then yields
an equilibrium temperature that decreases with increasing density.  The
work of \citet{KOY00}, which assumes that photoelectric heating
dominates, rather than cosmic ray heating as had been assumed in earlier
work, predicts a very similar trend of decreasing temperature with
increasing density at low densities.  The resulting temperature-density
relation can be approximated by a power law with an exponent of about
$-0.275$, which corresponds to a polytropic equation of state with
$\gamma = 0.725$.  The observational results of \citet{MYE78} shown in Fig.~2
of \citet{LAR85} suggest temperatures rising again toward the high end of
this low-density regime, but those measurements refer mainly to relatively
massive and warm cloud cores and not to the small, dense, cold cores in
which low-mass stars form.  As reviewed by \citet{EVA99}, the temperatures
of these cores are typically only about $8.5\,$K at a density of
$10^{-19}\,\mathrm{g\,cm}^{-3}$, consistent with a continuation of the decreasing trend noted
above and with the continuing validity of a polytropic approximation with
$\gamma \approx 0.725$ up to a density of at least $10^{-19}\,\mathrm{g\,cm}^{-3}$.

At densities higher than this, star-forming cloud cores become opaque
to the heating and cooling radiation that determines their temperatures at
lower densities, and at densities above $10^{-18}\,\mathrm{g\,cm}^{-3}$ the gas becomes
thermally coupled to the dust grains, which then control the temperature by
their far-infrared thermal emission.  In this high-density regime, dominated
thermally by the dust, there are few direct temperature measurements because
the molecules normally observed freeze out onto the dust grains, but most of
the available theoretical predictions are in good agreement concerning the
expected thermal behavior of the gas \citep{LAR73, LOW76, MAS00}. 
The balance between compressional heating and
thermal cooling by dust results in a temperature that increases slowly with
increasing density, and the resulting temperature-density relation can be
approximated by a power law with an exponent of about $0.075$, which
corresponds to $\gamma = 1.075$. Taking these values, the temperature is
predicted to reach a minimum of $5\,$K at the transition between the
low-density and the high-density regime at about $2 \times
10^{-18}\,\mathrm{g\,cm}^{-3}$, at which point the Jeans
mass is about $0.3\,M_{\odot}$ \citep[see also,][]{LAR05}. The actual minimum temperature reached is somewhat uncertain
because observations have not yet confirmed the predicted very low values,
but such cold gas would be very difficult to observe; various efforts to
model the observations have suggested central temperatures between $6\,$K and
$10\,$K for the densest observed prestellar cores, whose peak densities may
approach $10^{-17}\,\mathrm{g\,cm}^{-3}$ \citep[e.g.,][]{ZUC01, EVA01, TAF04}.  A power-law approximation to the equation of state
with $\gamma \approx 1.075$ is expected to remain valid up to a density of about
$10^{-13}\,\mathrm{g\,cm}^{-3}$, above which increasing opacity to the thermal emission
from the dust causes the temperature to begin rising much more rapidly,
resulting in an "opacity limit" on fragmentation that is somewhat below
$0.01\,M_{\odot}$ \citep{LOW76, MAS00}.

\section{Analytical Approach}
\label{sec:ana-app}
Following the above considerations, we use a polytropic equation of state to
describe the thermal state of the gas in our models with a polytropic exponent that changes
at a certain critical density $\rho_{\mathrm{c}}$ from $\gamma_1$ to $\gamma_2$:
\begin{eqnarray}
P & = & K_1\,\rho^{\gamma_1}~~~~~~~~~~\rho\le\rho_{\mathrm{c}}\nonumber \\
P & = & K_2\,\rho^{\gamma_2}~~~~~~~~~~\rho>\rho_{\mathrm{c}}
\label{equ:poly}
\end{eqnarray}
where $K_1$ and $K_2$ are constants, and P, and $\rho$  are thermal pressure
and gas density.
For an ideal gas, the equation of state is:
\begin{equation}
P=\frac{k_{\mathrm{B}}}{\mu m_{\mathrm{p}}}\rho T(\rho)
\end{equation}
where T is the temperature, and $k_{\mathrm{B}}$, $\mu$, and $m_{\mathrm{p}}$ are Boltzmann constant,
molecular weight, and proton mass.
So the constant K can be written as:
\begin{equation}
K=\frac{k_{\mathrm{B}}}{\mu m_{\mathrm{p}}}\rho^{1-\gamma}T(\rho)
\end{equation}
Since K is defined as a constant in $\rho$, it follows for T:
\begin{eqnarray}
T_1 & = & a_1\,\rho^{\gamma_1-1}~~~~~~~~~~\rho\le\rho_{\mathrm{c}}\nonumber \\
T_2 & = & a_2\,\rho^{\gamma_2-1}~~~~~~~~~~\rho>\rho_{\mathrm{c}}
\end{eqnarray}
where $a_1$ and $a_2$ are constants.
The initial conditions define $a_1$:
\begin{equation}
a_1=T_0\rho_0^{1-\gamma_1}
\end{equation}
At $\rho_{\mathrm{c}}$ it holds that:
\begin{equation}
T_1(\rho_{\mathrm{c}})=T_2(\rho_{\mathrm{c}})
\end{equation} 
Thus, $a_2$ can be written in terms of $a_1$:
\begin{equation}
a_2=a_1\rho_{\mathrm{c}}^{\gamma_1-\gamma_2}
\end{equation}

According to the analytical work by \citet{JEA02} on the stability of a
self-gravitating, isothermal medium the oscillation frequency $\omega$ and the
wavenumber $k$ of small perturbations satisfy the dispersion relation
\begin{equation}
\omega^2 - c_{\mathrm{s}}^2 k^2 +4 \pi G \rho_0 = 0 \,\,,
\end{equation}
where $c_{\mathrm{s}}$ is the sound speed, $G$ the gravitational
constant, and $\rho_0$ the gas density. The perturbation is unstable if the
wavelength~$\lambda$ exceeds the Jeans length
$\lambda_{\mathrm{J}}=2\pi/k_{\mathrm{J}}$ or, equivalently, if the mass
exceeds the Jeans mass
\begin{equation}
M_{\mathrm{J}}=\frac{4\pi}{3}\rho_0\left(\frac{\lambda_{\mathrm{J}}}{2}\right)^3=\frac{\pi^{5/2}}{6}G^{-3/2}{\rho_0}^{-1/2}c_{\mathrm{s}}^3\,\,.
\end{equation} 
Note that we define the Jeans mass~$M_{\mathrm{J}}$ as the mass originally
contained within a sphere of diameter~$\lambda_{\mathrm{J}}$.

\begin{figure}[t]
  \resizebox{\hsize}{!}{\includegraphics{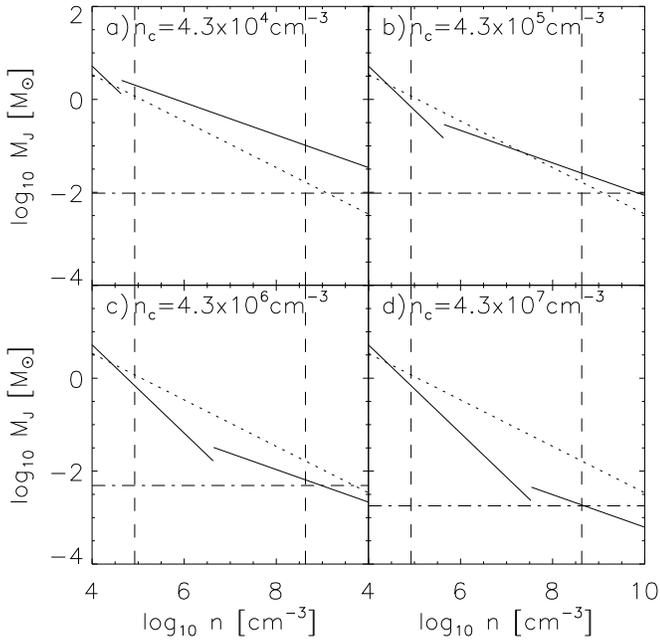}}
\caption{Local Jeans mass as a function of density for four runs with
  different critical densities $n_{\mathrm{c}}$. For comparison the dependence is also
  shown for the isothermal case (\textit{dotted line\nocorr}). The Jeans mass
  changes at the critical density. The initial mean density and the density at
  which sink particles form are represented by the vertical
  \textit{dashed\nocorr} lines. The \textit{dashed-dotted\nocorr} lines show
  the minimal resolvable mass for the runs with the highest resolution.}
\label{fig:all-jeans}
\end{figure}

In a system with a polytropic EOS, i.e., $P=K\rho^{\gamma}$, the sound speed is
\begin{equation}
c_s = \left(\frac{\mathrm{d}P}{\mathrm{d}\rho}\right)^{1/2}=(K\gamma)^{1/2}\rho^{(\gamma-1)/2}\,\,.
\end{equation} 
Thus, the Jeans mass can be written as
\begin{equation}
M_\mathrm{J}=\frac{\pi^{5/2}}{6}\left(\frac{K}{G}\right)^{3/2}\gamma^{3/2}\rho^{(3/2)\gamma-2}\,\,.
\label{equ:mj}
\end{equation}
Using Eqs.~3, 4, 5, 7 and 11 one finds:
\begin{eqnarray*}
M_{\mathrm{J}1}& = & \frac{\pi^{5/2}}{6}\left(\frac{k_{\mathrm{B}}T_0\rho_0^{1-\gamma_1}}{G\mu m_{\mathrm{p}}}\right)^{3/2}\gamma_1^{3/2}\rho^{(3/2)\gamma_1-2}~~~~~~~~~~~~~~~~~~~\rho\le\rho_{\mathrm{c}}\nonumber \\
M_{\mathrm{J}2}& = & \frac{\pi^{5/2}}{6}\left(\frac{k_{\mathrm{B}}T_0\rho_0^{1-\gamma_1}}{G\mu m_{\mathrm{p}}}\right)^{3/2}\gamma_2^{3/2}\rho_{\mathrm{c}}^{(3/2)(\gamma_1-\gamma_2)}\rho^{(3/2)\gamma_2-2}~~\rho>\rho_{\mathrm{c}}
\end{eqnarray*} 
The sound speed changes when the polytropic index changes at $\rho_{\mathrm{c}}$,
so $M_{\mathrm{J}}$ also varies (see
Fig.~\ref{fig:all-jeans}), such that:

\begin{equation}
\frac{M_{\mathrm{J}1}}{M_{\mathrm{J}2}}=\left(\frac{\gamma_1}{\gamma_2}\right)^\frac{3}{2}.
\end{equation}
If we use $\gamma_1$=0.7 and
$\gamma_2$=1.1 as justified in Sect.~\ref{sec:therm-prop} then
$M_{\mathrm{J}1}\propto \rho^{-0.95}$. 

 During the initial phase of collapse, the turbulent flow produces strong
  ram pressure gradients that form density enhancements. Higher density leads to
smaller local Jeans masses, so these regions begin to collapse and fragment.
Simulations with an SPH code different from the one used in the present work
show that fragmentation occurs more efficiently for smaller values of $\gamma$,
and less efficiently for $\gamma > 1$, cutting off entirely at $\gamma > 1.4$
\citep{LI03, ARC91}. For filamentary systems, fragmentation already stops for
$\gamma > 1$ \citep{KAW98}. This point is discussed in more detail in Sec.~\ref{sec:res}.

What happens when $\gamma$ increases above unity at the critical density
$\rho_{\mathrm{c}}$?  One suggestion is that the increase in $\gamma$ is sufficient to
strongly reduce fragmentation at higher densities, introducing a characteristic
scale into the mass spectrum at the value of the Jeans mass at
$\rho_{\mathrm{c}}$. Then
the behavior of the Jeans mass with increasing critical density would
immediately allow us to derive the scaling law
\begin{equation}
 M_{\mathrm{ch}}\propto\rho_{\mathrm{c}}^{-0.95}.
\label{eq:scale}
\end{equation}

This simple analytical consideration would then predict a
characteristic mass scale which corresponds to a peak of the IMF at 0.35 $M_{\odot}$ for a
critical density of $\rho_{\mathrm{c}}=10^{-18}\,\mathrm{g\, cm^{-3}}$ or
equivalently a number density of $n_{\mathrm{c}}=2.5\times10^5\,\mathrm{cm^{-3}}$
when using a mean molecular weight $\mu=2.36$ appropriate for solar
metallicity molecular clouds in the Milky Way. Note, however, that this
simple scaling law does not take any further dynamical processes into account.

\section{Numerical Method}
\label{sec:num}
\begin{table}[t]
\caption{Sample parameters, name of the environment used in the text, driving
  scale $k$, critical density $n_{\mathrm{c}}$, number of SPH particles,
  number $\mathcal{N}$ of protostellar objects (i.e.,\ ``sink particles'' in
  the centers of protostellar cores) at final stage
  of the simulation, percentage of accreted mass at final stage $
M_{\mathrm{acc}}/M_{\mathrm{tot}}$} 
\begin{tabular}{lrrrrl} \hline \hline
Name & $k$ & log$_{10}$ $n_{\mathrm{c}}$ & particle & $\mathcal{N}$ & $
\frac{M_{\mathrm{acc}}}{M_{\mathrm{tot}}}$ \\
& & $\left[\mathrm{cm^{-3}}\right]$ & number &
&$\left[\%\right]$ \\\hline
Ik2 & 1..2 & --- & $205\,379$ & 59 & 56\\
Ik2b & 1..2 & --- & $1\,000\,000$ & 73 & 78\\
Ik2L & 1..2 & --- & $9\,938\,375$ &  6 & 4 \\
R5k2 & 1..2 & $4.63$  & $205\,379$ & 22 & 73\\
R5k2b & 1..2 & $4.63$  & $1\,000\,000$ & 22 & 70\\
R6k2 & 1..2 & $5.63$  & $205\,379$ & 64 & 93 \\
R6k2b & 1..2 & $5.63$  & $1\,000\,000$ & 54 & 61 \\
R7k2 & 1..2 & $6.63$  & $205\,379$ & 122 & 84 \\
R7k2b & 1..2 & $6.63$  & $1\,000\,000$ & 131 & 72 \\
R7k2L & 1..2 & $6.63$  & $1\,953\,125$ & 143 & 46 \\
R8k2 & 1..2 & $7.63$  & $205\,379$ & 194 & 78 \\
R8k2b & 1..2 & $7.63$  & $1\,000\,000$ & 234 & 53 \\
R8k2L & 1..2 & $7.63$  & $5\,177\,717$ & 309 & 29 \\ 
R5k8 & 7..8 & $4.63$  & $205\,379$ & 1 & 64\\
R6k8 & 7..8 & $5.63$  & $205\,379$ & 38 & 68 \\
R7k8 & 7..8 & $6.63$  & $205\,379$ & 99 & 60 \\
R8k8 & 7..8 & $7.63$  & $205\,379$ & 118 & 72 \\
R5k2r1 & 1..2 & $4.63$  & $205\,379$ & 16 & 62\\
R6k2r1 & 1..2 & $5.63$  & $205\,379$ & 34 & 72 \\
R7k2r1 & 1..2 & $6.63$  & $205\,379$ & 111 & 68 \\
R8k2r1 & 1..2 & $7.63$  & $205\,379$ & 149 & 64 \\
R5k2r2 & 1..2 & $4.63$  & $205\,379$ & 21 & 72\\
R6k2r2 & 1..2 & $5.63$  & $205\,379$ & 51 & 74 \\
R7k2r2 & 1..2 & $6.63$  & $205\,379$ & 119 & 70 \\
R8k2r2 & 1..2 & $7.63$  & $205\,379$ & 184 & 70 \\
R5k2r3 & 1..2 & $4.63$  & $205\,379$ & 18 & 90\\
R6k2r3 & 1..2 & $5.63$  & $205\,379$ & 52 & 85 \\
R7k2r3 & 1..2 & $6.63$  & $205\,379$ & 123 & 76 \\
R8k2r3 & 1..2 & $7.63$  & $205\,379$ & 196 & 71 \\\hline
\end{tabular}
\label{tab:prop}
\end{table}
\begin{figure}
  \resizebox{\hsize}{!}{\includegraphics{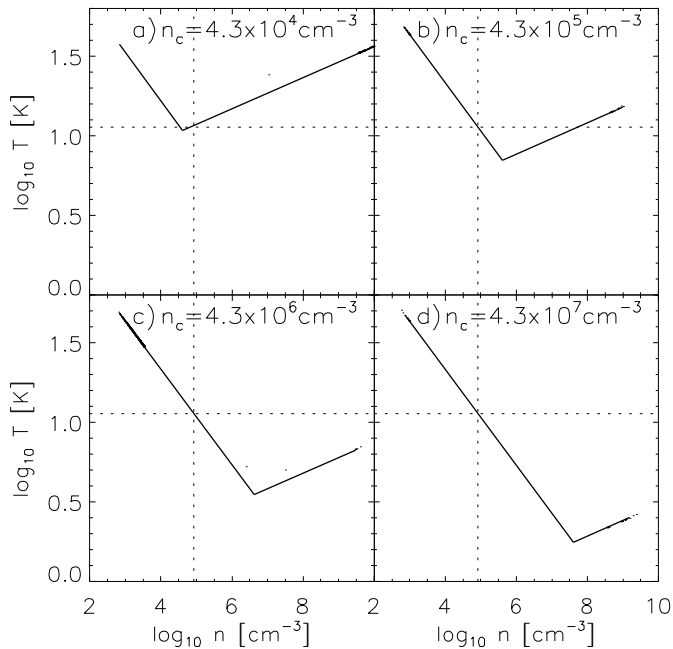}}
\caption{Temperature as a function of density for four runs with
  different critical densities $n_{\mathrm{c}}$. The {\it
    dotted lines \nocorr} show the initial conditions. The curve has a
    discontinuous derivative at the critical density.}
\label{fig:all-temprho}
\end{figure}
 In order to test the scaling of the characteristic mass~$M_{\mathrm{ch}}$
  given by Eq.~\ref{eq:scale}, we carry out simulations of regions in
  turbulent molecular clouds in which we vary the critical
  density~$\rho_{\mathrm{c}}$ and determine the resulting mass spectra of
  protostellar objects, and thus $M_{\mathrm{ch}}$.
During gravoturbulent
fragmentation it is necessary to follow the gas over several orders of
magnitude in density. The method of choice therefore is smoothed
particle hydrodynamics (SPH). Excellent overviews of the
method, its numerical implementation, and some of its applications are
given in reviews by \citet{BEN90} and \citet{MON92}. We use the parallel code GADGET,
designed by \citet{SPR01}. SPH is a Lagrangian method, where the fluid
is represented by an ensemble of particles, and flow quantities are
obtained by averaging over an appropriate subset of SPH particles.  We
  use a spherically symmetric cubic spline function to define the smoothing
  kernel \citep{MON92}. The
  smoothing length can vary in space and time, such that the number of
  considered neighbors is always approximately 40. The
method is able to resolve large density contrasts as particles are
free to move, and so naturally the particle concentration increases in
high-density regions. We use the \citet{BAT97} criterion to determine the
resolution limit of our calculations. It is adequate for the problem
considered here, where we follow the evolution of highly nonlinear
density fluctuations created by supersonic turbulence. We have
performed a resolution study with up to $10^7$ SPH particles to
confirm this result.

\subsection{Sink particles}
SPH simulations of collapsing regions become slower as more particles move to
higher density regions and hence have small timesteps. Replacing dense cores
by one single particle leads to considerable increase of the overall
computational performance.
 Introducing sink particles allows us to
follow the dynamical evolution of the system over many free-fall times. 
  Once the density contrast in the center of a collapsing cloud core exceeds a
  value of 5000, corresponding to a threshold density of about
  $n_{\mathrm{th}}=4\times10^8\,\mathrm{cm^{-3}}$ (see Fig.~\ref{fig:all-jeans}), the entire central region is replaced by a ``sink particle''
  \citep*{BAT95}. It is a single, non-gaseous, massive
particle that only interacts with normal SPH particles via gravity. Gas particles that come within a certain radius of the sink
particle, the accretion radius~$r_{\mathrm{acc}}$, are accreted if they are
bound to the sink particle. This allows us to keep track of the total mass,
the linear and angular momentum of the collapsing gas. 

Each sink particle
defines a control volume with a fixed radius of 310~AU. This radius is chosen
such that we always resolve the Jeans scale below the threshold density
$n_{\mathrm{th}}$, following \citet{BAT97}.  
We
cannot resolve the subsequent evolution in its interior. 
Combination
with a detailed one-dimensional implicit radiation hydrodynamic method shows that a
protostar forms in the very center about $10^3\,\mathrm{yr}$ after sink
creation \citep{WUC01}. We subsequently call the sink 
{\it protostellar object\nocorr} or simply {\it protostar\nocorr}.
Altogether, the sink particle represents only the innermost, highest-density
part of a larger collapsing region. The technical details on the
implementation of sink particles in the parallel SPH code GADGET can be found
in Appendix~A.  




Protostellar collapse is 
accompanied by a substantial loss of specific angular momentum, even in the
absence of magnetic fields \citep{JAP04}. Still, most of the matter that falls in will assemble in
a protostellar disk. It is then transported inward
by viscous and possibly gravitational torques
\citep[e.g.,][]{BOD95,PAP95,LIN96}.  With typical disk sizes of order
of several hundred AU, the control volume fully encloses both star
and disk. If low angular momentum material is accreted, the disk is
stable and most of the material ends up in the central star. In this
case, the disk simply acts as a buffer and smooths eventual accretion
spikes. It will not delay or prevent the mass growth of the central
star by much.  However, if material that falls into the control
volume carries large specific angular momentum, then the mass load
onto the disk is likely to be faster than inward transport. The disk
grows large and may become gravitationally unstable and fragment.
This may lead to the formation of a binary or higher-order multiple
\citep{BOD00, FRO02}.
\subsection{Model parameters}
\begin{figure*}[t]
\centering
  \includegraphics[width=14cm]{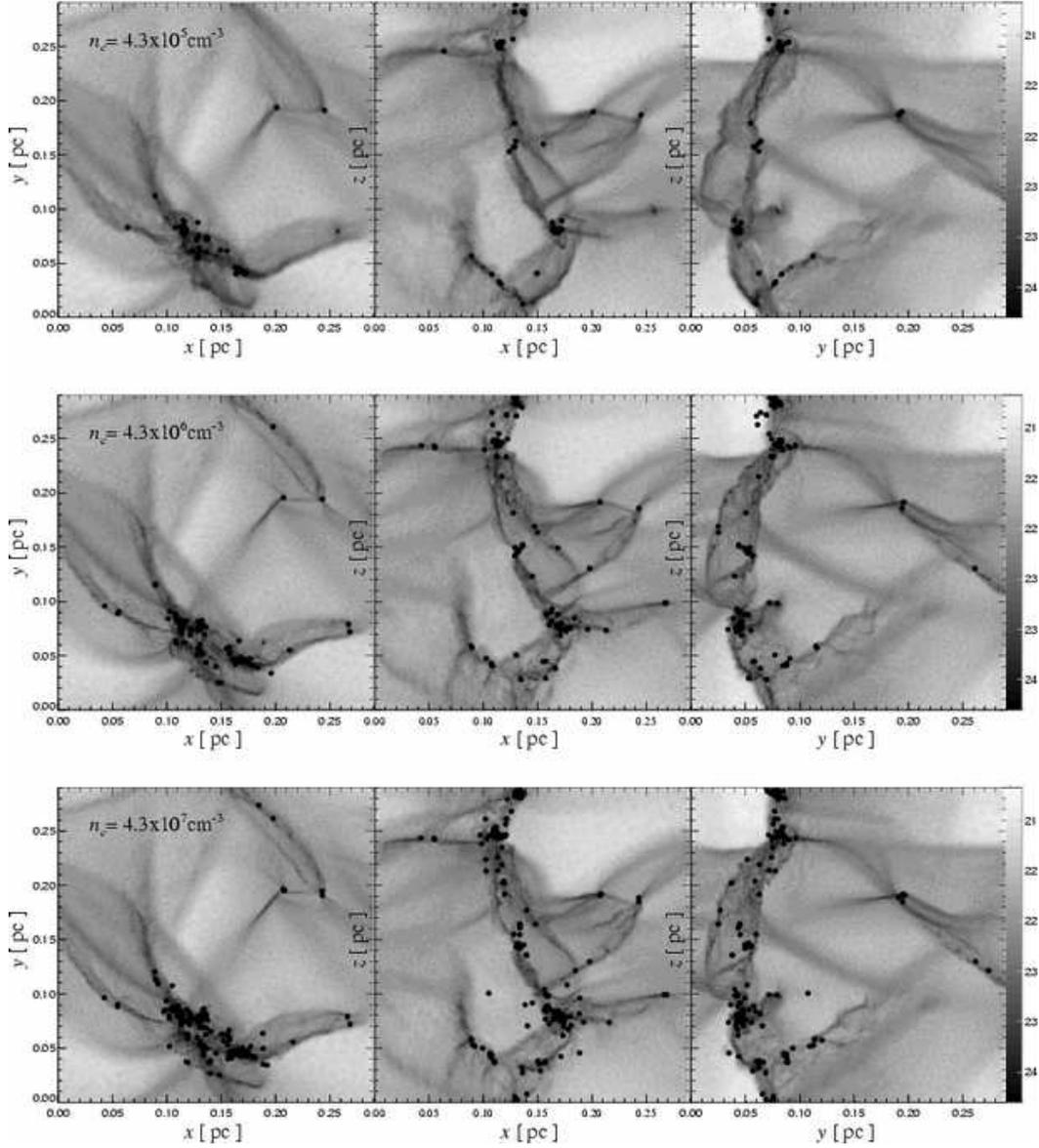}
\caption{Column density distribution of the gas and location of identified
  protostellar objects (\textit{black circles\nocorr}) using the high-resolution models R6..8k2b at the stage where approximately 50$\%$ of the gas is
  accreted. Projections in the xy-, xz-, and yz-plane are shown for three
  different critical densities.}
\label{fig:all-density}
\end{figure*} 
We include turbulence in our version of the code that is driven uniformly with the method described by
\citet{MAC98} and \citet{MAC99}.  The observed turbulent velocity field in
  molecular clouds will decay in a crossing time if not continuously
  replenished \citep{MAC04,
  SCA04, ELM04}. If the turbulence decays, then only thermal pressure prevents
global collapse. In our case, we examine regions globally supported by the
turbulence at the initial time. We choose to model continuously driven turbulence
leading to inefficient star formation rather than a globally
collapsing region producing efficient star formation \citep{MAC04}. This is achieved
here by applying a nonlocal driving scheme that inserts energy in a limited
range of wave numbers~$k$. \citet{MAC99} shows that hydrodynamical turbulence
decays with a constant energy-dissipation coefficient. Thus a constant kinetic
energy input rate~$\dot{E}_{\mathrm{in}}$ is able to maintain the observed turbulence
and to stabilize the system against gravitational contraction on
global scales. The driving strength is adjusted to yield a constant
turbulent rms Mach number~$\mathcal{M}_{\mathrm{rms}}=3.2$. Furthermore, it is
known that the velocity structure in molecular clouds is always dominated by
the largest-scale mode \citep[e.g.,][]{MAC00,OSS01,OSS02,BRU04a}, consequently we insert energy on scales of
the order of the size of our computational domain, i.e. with wave numbers
$k=1..2$. In the adopted integration scheme, we add the turbulent energy
$\Delta E_{\mathrm{in}}=\dot{E}_{\mathrm{in}}\Delta t$ to the system at each timestep $\Delta t$
whenever more than $40\%$ of the SPH particles are advanced.

In all our models we adopt an initial temperature of
$11.4\,\mathrm{K}$ corresponding to a sound speed $c_s
=0.2\,\mathrm{km\,s^{-1}}$, a molecular weight $\mu$ of 2.36 and an initial
number density of
$n=8.4 \times 10^4\,\mathrm{cm^{-3}}$, which is typical for
star-forming molecular cloud regions (e.g., $\rho$-Ophiuchi, see Motte et al. 1998
 or the central region of the Orion Nebula
Cluster, see Hillenbrand 1997; Hillenbrand \& Hartmann 1998). 
\nocite{MOT98,HIL97,HIL98}
Our simulation cube holds a mass of $120\,M_{\odot}$
and has a size of $L=0.29\,\mathrm{pc}$. The cube is subject to
periodic boundary conditions in every direction. The mean initial Jeans mass is $\langle
M_{\rm J} \rangle _{\mathrm{i}} = 0.7\,M_{\odot}$. 

We use the EOS described in Sect.~\ref{sec:ana-app}, and compute models with $4.3\times10^4\,\mathrm{cm^{-3}} \le n_{\mathrm{c}} \le
4.3\times10^7\,\mathrm{cm^{-3}}$. 
 Note, that the lowest and the highest of these critical densities represent rather
  extreme cases. From Fig.~\ref{fig:all-temprho}, where we show the temperature as a
function of number density, it is evident that they result in temperatures that
are too high or too low compared to observations and theoretical
predictions. Nevertheless, including these
cases clarifies the existence of a real trend in the dependence of the
characteristic mass scale on the critical density.
Each simulation starts with a uniform
density. Driving begins immediately, while self-gravity is turned on at $t=2.0\,t_{\mathrm{ff}}$, after turbulence is fully established. The global free-fall timescale is
$t_{\mathrm{ff}}\approx10^5\,\mathrm{yr}$.
Our models are named mnemonically. R5 up to R8 stand for the critical density~$n_\mathrm{c}$ ($4.3\times10^4\,\mathrm{cm^{-3}} \le n_{\mathrm{c}} \le
4.3\times10^7\,\mathrm{cm^{-3}}$) in the equation of state, k2 or k8 stand for
the wave numbers ($k=1..2$ or $k=7..8$) at which the driving energy is
injected into the system and b flags the runs with 1 million gas
particles. The letter L marks the high resolution runs for critical densities
$n_{\mathrm{c}} = 4.3\times10^6\,\mathrm{cm^{-3}}$ and $n_{\mathrm{c}} =
4.3\times10^7\,\mathrm{cm^{-3}}$ with 2 million and 5.2 million particles, respectively. Different realizations of the
  turbulent velocity field are denoted by r1, r2, r3. For comparison we also
  run isothermal simulations marked with the letter I that have particle
  numbers of approximately 200000, 1 million and 10 million gas particles. 

The number of
particles determines the minimal resolvable Jeans mass in our
models. Figure~\ref{fig:all-jeans} shows the dependence of the local Jeans
mass on the density for the adopted polytropic equation of state. 
At the critical density the dependency of the Jeans mass on density changes
its behavior. The minimum Jeans mass~$M_{\mathrm{res}}$ that needs to be resolved
occurs at the density at which sink particles are formed. 
A local Jeans mass is
considered resolved if it contains at least $2\times N_{\mathrm{neigh}}= 80$
SPH particles \citep{BAT97}. As can be seen in Fig.~\ref{fig:all-jeans} we
are able to resolve $M_{\mathrm{res}}$ with 1 million particles for critical
densities up to $n_{\mathrm{c}}= 4.3\times10^5\,\mathrm{cm^{-3}}$. Since this
is not the case for $n_{\mathrm{c}}= 4.3\times10^6\,\mathrm{cm^{-3}}$ and $n_{\mathrm{c}}= 4.3\times10^7\,\mathrm{cm^{-3}}$, we
repeat these simulations with 2 million and 5.2 million particles, respectively. 
Due to long calculation times we follow the latter only to the point in time where about $30\%$ of the gas has been accreted.   

At the density where $\gamma$ changes from below unity to above unity, the
temperature reaches a minimum. This is reflected in the "V"-shape shown in
Figure~\ref{fig:all-temprho}. 
All our simulations start with the same initial conditions in temperature and
density as marked by the dotted lines.  
   
In a further set of simulations we analyze the influence of changing the turbulent driving scheme on fragmentation while using a polytropic equation of
state. These models contain $2\times10^5$ particles each. 

Following \citet{BAT97}, these runs are not considered fully resolved for
$n_{\mathrm{c}}\geq4.3\times 10^5$ at the density of sink particle creation,
since $M_{\mathrm{res}}$ falls below the critical mass of 80~SPH particles. We note,
however, that the global accretion history is not strongly affected. 
First, we study the effect of different realizations of the turbulent driving
fields on typical masses of protostellar objects. We simply select different random numbers to generate the field while
keeping the overall statistical properties the same. This allows us to assess
the statistical reliability of our results. These models are labeled from
R5..8k2r1 to R5..8k2r3.
Second, driving in two different wavenumber ranges is considered. Most models are driven
on large scales ($1\le k\le2$) but we have run a set of models driven on small
scales ($7\le k\le8$) for comparison.


The main model parameters are summarized in Table~\ref{tab:prop}. 
\section{Gravoturbulent fragmentation in polytropic gas}
\label{sec:res}
\begin{figure}[t]
  \resizebox{\hsize}{!}{\includegraphics{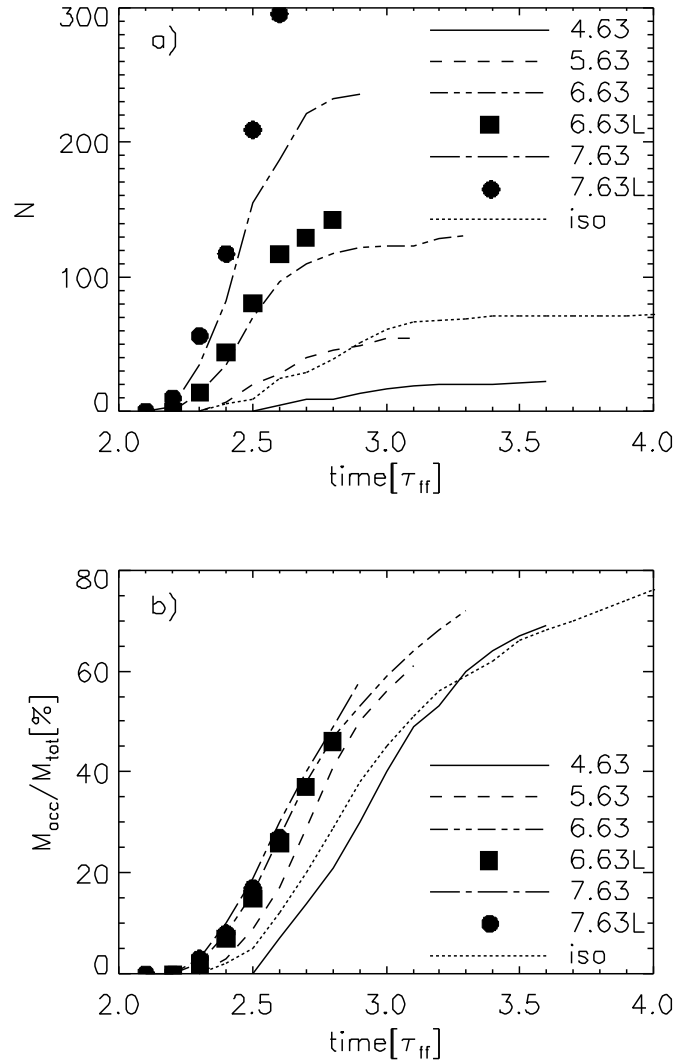}}
\caption{Temporal evolution of the number of protostellar objects ({\it upper plot
  \nocorr}) and of the ratio of accreted gas mass to total gas mass ({\it lower plot
  \nocorr}) for models~R5..8k2b. The legend shows the logarithms of the
  respective number densities in cm$^{-3}$. The times are given in units of a
  free-fall time~$\tau_{\mathrm{ff}}$. We also show the models R7k2L and R8k2L
  with $2\times10^6$ and $5\times10^6$ particles, respectively, which are
  denoted by the letter ``L''.  For comparison the {\it dotted lines}
  indicate the values for the
  isothermal model~Ik2b.}
\label{fig:all-time}
\end{figure}

Turbulence establishes a complex network of interacting shocks,
where converging flows and shear generate filaments of high density. 
The interplay between gravity and thermal pressure determines the further
dynamics of the gas. Adopting a polytropic EOS (Eq.~\ref{equ:poly}), the
choice of the polytropic exponent plays an important role determining the
fragmentation behavior. From Eq.~\ref{equ:mj} it is evident that
$\gamma=4/3$ constitutes a critical value. A Jeans mass analysis shows that
for three-dimensional structures
$M_\mathrm{J}$ increases with increasing density if $\gamma$ is above
$4/3$. Thus, $\gamma > 4/3$ results in the termination of any gravitational
collapse.

Also, collapse and fragmentation in filaments depend on the
equation of state. The equilibrium and stability of filamentary structures has been studied extensively,
beginning with \cite{CHA53}, and this work has been reviewed
by \citet{LAR85, LAR03}.  For many types of collapse problems, insight into
the qualitative behavior of a collapsing configuration can be gained from
similarity solutions \citep{LAR03}. For the collapse
of cylinders with an assumed polytropic equation of state solutions have been derived
by \citet{KAW98}, and these authors found that the existence of
such solutions depends on the assumed value of $\gamma$: similarity solutions
exist for $\gamma < 1$ but not for $\gamma > 1$.  These authors also found that
for $\gamma < 1$, the collapse becomes slower and slower as
$\gamma$ approaches unity from below, asymptotically coming to a halt when
$\gamma = 1$.  This result shows in a particularly clear way that $\gamma = 1$
is a critical case for the collapse of filaments. \citet{KAW98}
suggested that the slow collapse that is predicted to occur for $\gamma$
approaching unity will in reality cause a filament to fragment into clumps,
because the timescale for fragmentation then becomes shorter than the
timescale for collapse toward the axis of an ideal filament.  If the effective value of $\gamma$
increases with increasing density as the collapse proceeds, as is expected
from the predicted thermal behavior discussed in Sect.~\ref{sec:therm-prop}, fragmentation
may then be particularly favored to occur at the density where $\gamma$
approaches unity. In their numerical study \citet{LI03} found, for a range of assumed polytropic
equations of state, that the amount of fragmentation that occurs is indeed very
sensitive to the value of the polytropic exponent $\gamma$, especially for
values of $\gamma$ near unity \citep[see also,][]{ARC91}. 

The fact that filamentary structure is so prominent in our results and
other simulations of star formation, together with the fact that most of
the stars in these simulations form in filaments, suggests that the
formation and fragmentation of filaments may be an important mode of star
formation quite generally.  This is also supported by the fact that many
observed star-forming clouds have filamentary structure, and by the
evidence that much of the star formation in these clouds occurs in filaments
\citep{SCH79, LAR85, CUR02, HAR02}.
As we note in Sects.~\ref{sec:therm-prop} and \ref{sec:ana-app} and following
\citet{LAR73, LAR05}, the Jeans mass at the density where the temperature
reaches a minimum (see Fig.~\ref{fig:all-temprho}), and hence, $\gamma$ approaches unity, is
predicted to be about $0.3\,M_{\odot}$, coincidentally close to the mass
at which the stellar IMF peaks. This similarity is an additional hint that filament fragmentation with a
varying polytropic exponent may play an important role in the origin of
the stellar IMF and the characteristic stellar mass.

The filamentary structure that occurs in our simulations is visualized in Fig.~\ref{fig:all-density}.
Here we show the column density distribution of the gas and the
distribution of protostellar objects. We display the results for three different
critical densities in   xy-, xz- and yz-projection. The volume density is computed
from the SPH kernel in 3D and then projected along the three principal axes. 
Figure~\ref{fig:all-density} shows for all three cases a remarkably filamentary
structure. These filaments define the loci where most protostellar objects
form. 

Clearly, the change of the polytropic exponent~$\gamma$ at a certain critical
density influences the number of protostellar objects. If the critical density
increases then more protostellar
objects form but the mean mass decreases.

We show this quantitatively in Fig.~\ref{fig:all-time}.
In (a) we compare the number of protostellar objects for
different critical densities~$n_{\mathrm{c}}$ for the models~R5...8k2b. The rate at which
new protostars form changes with different $n_{\mathrm{c}}$. Models that switch
from low $\gamma$ to high $\gamma$ at low densities built up protostellar objects
more rarely than models that change $\gamma$ at higher densities. 

Figure~\ref{fig:all-time}b shows the accretion histories (the time evolution of the combined
mass fraction of all protostellar objects) for the models~R5..8k2b. Accretion
starts for all but one case approximately at the same time. In model R5k2b,
$\gamma = 1.1$ already at the mean initial density, thus $\gamma$ does not change during collapse. In this case accretion starts at a later time. This confirms the finding by
\citet{LI03} that accretion is delayed for large $\gamma$. In the other four
cases the accretion history is very similar and the slope is approximately the
same for all models. 

In both plots we also show the results from our high resolution runs R7k2L and
R8k2L. These simulations with 2 million and 5.2 million particles,
respectively, have an accretion history similar to the time evolution of the
accreted mass fraction in the runs with 1 million particles. The number of
protostellar objects, however, is larger for the runs with increased particle numbers.   
Combining our results in these two figures we find that an environment where
$\gamma$ changes at higher densities produces more, but less massive objects.
Thus, the mean mass of protostellar objects does indeed depend on the critical
density where $\gamma$ changes from 0.7 to 1.1.
\section{Dependence of the characteristic mass on the equation of state}
\label{sec:char}    
\begin{figure*}[t!]          
 \centering
  \includegraphics[width=16cm]{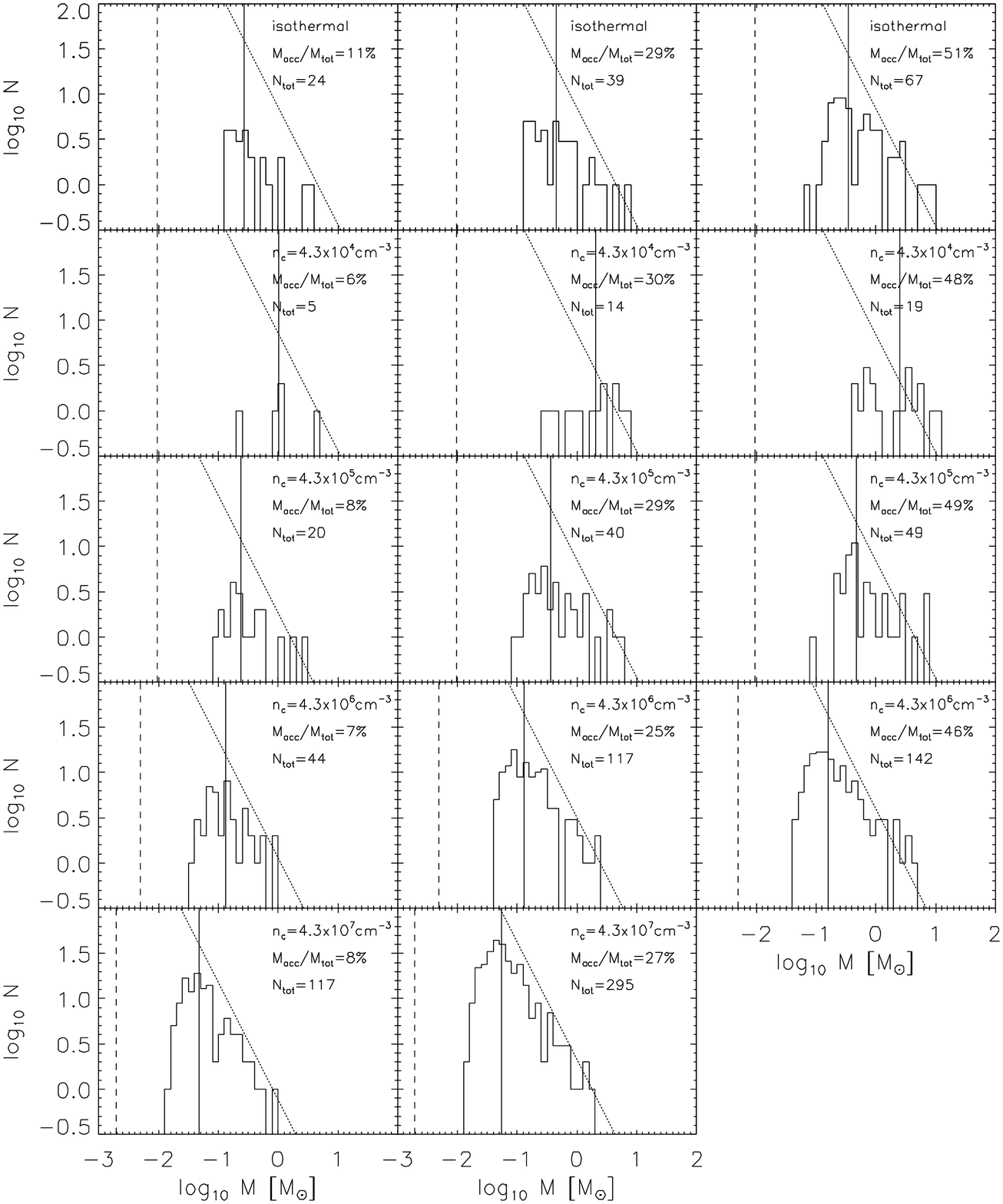}
\caption{Mass spectra of protostellar objects for models~R5..6k2b, model~R7k2L and model~R8k2L  at $10\%$,
  $30\%$ and $50\%$ of total mass accreted on these protostellar objects. For comparison we
  also show in the first row the mass spectra of the isothermal run~Ik2b. Critical
  density~$n_{\mathrm{c}}$, ratio of accreted gas mass to total gas mass
  $M_{\mathrm{acc}}/M_{\mathrm{tot}}$ and number of protostellar objects are given
  in the plots. The {\it vertical solid line \nocorr}shows the position of
  the median mass. The {\it dotted line \nocorr}has a slope of -1.3 and
  serves as a reference to the Salpeter value \citep{SAL55}. The {\it dashed
  line \nocorr}indicates the mass resolution limit.}
\label{fig:all-mspec}
\end{figure*}
 
Further insight into how the characteristic stellar mass may depend on the
critical density can be gained from the mass spectra of the protostellar
objects, which we show in Fig.~\ref{fig:all-mspec}. 
We plot the mass spectra
of models~R5...6k2b, model~R7k2L and model~R8k2L at different times, when the fraction of mass accumulated in protostellar
objects has reached approximately $10\%, 30\%$ and $50\%$. In the top row we
also display the results of an isothermal run for comparison. We used the same initial
conditions and parameters in all models shown. Dashed lines indicate the specific
mass resolution limits.

We find closest
correspondence with the observed IMF \citep[see, ][]{SCA98a, KRO02, CHA03} for a
critical density of $4.3\times10^6\,\mathrm{cm^{-3}}$ and for
stages of accretion around $30\%$ and above. At high masses, our distribution
follows a Salpeter-like power law. For comparison we indicate the Salpeter slope $x\approx1.3$
\citep{SAL55} where the IMF is defined by $\mathrm{d}N/\mathrm{d}\log m
\propto m^{-x}$. For masses about the median mass the distribution exhibits a
small plateau and then falls off towards smaller masses.

The model~R5k2b where the change in $\gamma$ occurs below the initial mean
density, shows a flat distribution with only few, but massive protostellar objects. They
reach masses up to $10\,M_{\odot}$ and the minimal mass is about
$0.3\,M_{\odot}$.
 All other models build up a power-law tail towards high masses. This is due to
  protostellar accretion processes, as more and more gas gets turned into
  stars \citep[see also,][]{BON01b,
  KLE01c, SCH04}. 
The distribution becomes more peaked for higher $n_{\mathrm{c}}$ and there is a
shift to lower masses. This is already visible in the mass
spectra when the protostellar objects have only accreted $10\%$ of the total
mass. Model~R8k2L has minimal and maximal masses of $0.013\,M_{\odot}$ and
$1.0\,M_{\odot}$, respectively. There is a gradual shift in the median mass
(as indicated by the vertical line) from
Model~R5k2b, with $M_{\mathrm{med}}=2.5\,M_{\odot}$ at 50\%, to Model~R8k2L,
with $M_{\mathrm{med}}=0.05\,M_{\odot}$ at 50\%. A similar trend is noticeable
during all phases of the model evolution.

\begin{figure}[t!]
  \resizebox{\hsize}{!}{\includegraphics{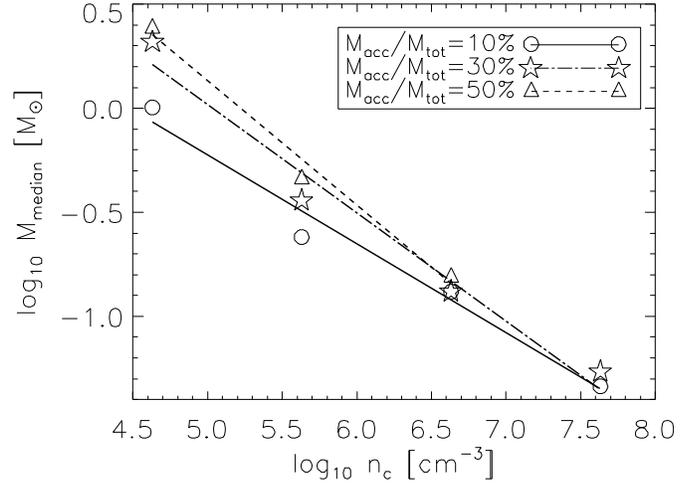}}
\caption{The plot shows the median mass of the protostellar objects over
  critical density for models~R5..6k2b, model~R7k2L and model~R8k2L. We display results for different
  ratios of accreted gas mass to total gas
  mass~$M_{\mathrm{acc}}/M_{\mathrm{tot}}$. We fit our data with straight lines for different
  stages of accretion. The slopes have the following values: $-0.43\pm0.05$ ({\it
  solid line\nocorr}), $-0.52\pm0.06$ ({\it
  dashed-dotted line\nocorr}), $-0.60\pm0.07$ ({\it
  dashed line\nocorr}).}
\label{fig:all-med-rho}
\end{figure}

This change of median mass with critical density~$n_{\mathrm{c}}$ is depicted in
Fig.~\ref{fig:all-med-rho}. Again, we consider models~R5...6k2b, model~R7k2L and model~R8k2L. The
median mass decreases clearly with increasing critical density as
expected. 
As we resolve
higher density contrasts the median collapse mass decreases. We fit our data
with a straight line. The slope takes values between $-0.4$ and $-0.6$.
These values are larger than the slope $-0.95$ derived from the simple scaling law (Eq.~\ref{eq:scale})
based on calculating the Jeans mass $M_\mathrm{J}$ at the critical density $n_\mathrm{c}$

One possible reason for this deviation is
the fact that most of the protostellar objects are members of tight groups. Hence, they are subject to mutual interactions and competitive
accretion that may change the environmental context for individual
protostars. This in turn influences the mass distribution and its
characteristics \citep[see, e.g., ][]{BON01a, BON01b}. 
 Another possible reason is that the mass that goes into filaments and then into collapse may depend on further
environmental parameters, some of which we discuss in Sect.~\ref{sec:dep}.

\section{Dependence of the characteristic mass on environmental
  parameters}
\label{sec:dep}


\subsection{Dependence on realization of the turbulent velocity field}
\begin{figure}[t!]
  \resizebox{7.2cm}{!}{\includegraphics[width=3cm]{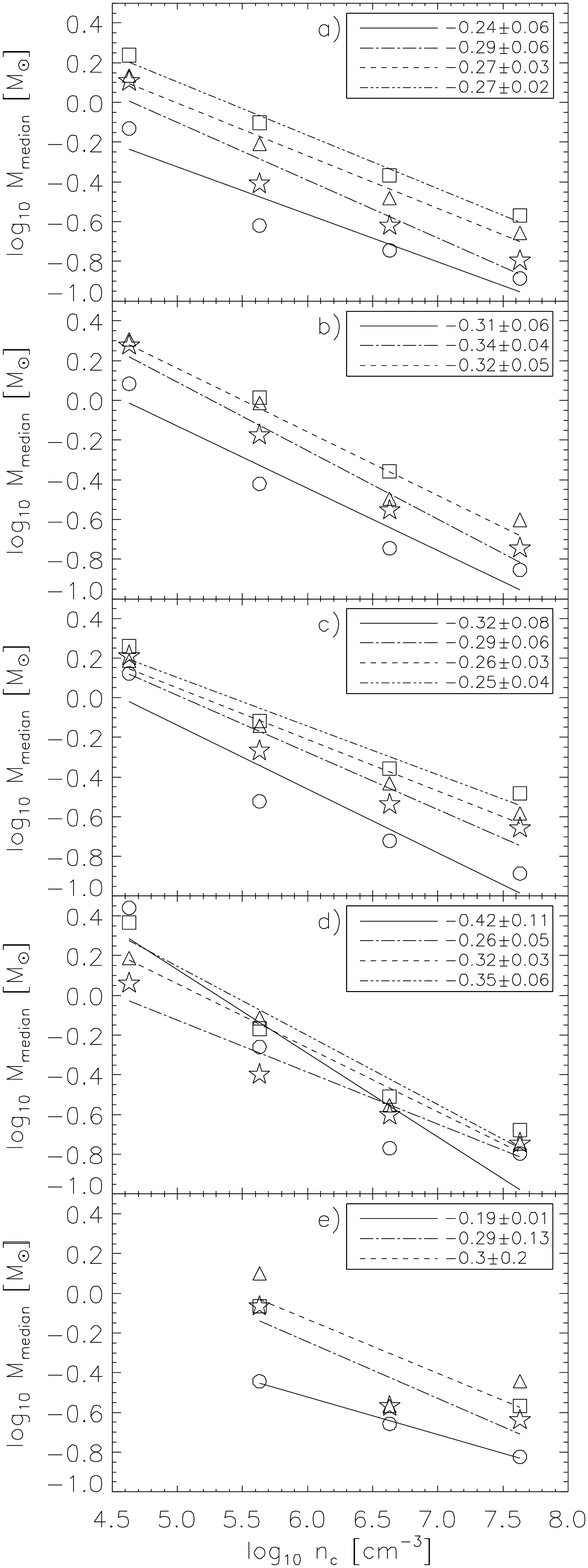}}
\caption{Median mass of protostellar objects over critical density at
  different evolutionary phases (ratio of accreted gas mass over total gas
  mass~$M_{\mathrm{acc}}/M_{\mathrm{tot}}$:10$\%$ ({\it
  circle\nocorr}), 30$\%$ ({\it
  star\nocorr}), 50$\%$ ({\it
  triangle\nocorr}), 70$\%$ ({\it
  square\nocorr})). In (a)-(d)
  we show identical models but with different turbulent velocity fields (same
  root-mean-square velocity). The model in (e) is driven on a smaller scale $k=7..8$ than in the other
  cases. All relevant parameters are summarized in Table~\ref{tab:prop}. We
  fit the median values with straight lines for different
  stages of accretion. The slopes are given in the plot and denoted as follows: 10$\%$ -- {\it
  solid line\nocorr}, 30$\%$ -- {\it
  dashed-dotted line\nocorr}, 50$\%$ -- {\it
  dashed line\nocorr}, 70$\%$ -- {\it
  dashed-double-dotted line\nocorr}.}
\label{fig:all-slope}
\end{figure}

We compare models with different realizations
of the turbulent driving field in Fig.~\ref{fig:all-slope}. We fit our data with straight lines for each
stage of accretion.
Figure~\ref{fig:all-slope}a shows the results of models~R5..8k2 which were
calculated with the same parameters but lower resolution than the models used
for Fig.~\ref{fig:all-med-rho}. As discussed in Sec.~\ref{sec:res}, although
the number of protostellar objects changes with the number of particles in the simulation, the time evolution of the total mass accreted on all
protostellar objects remains similar. Thus, lower resolution models exhibit the
same general trend as their high-resolution counterparts and show the same
global dependencies. We notice, however, that the slope of the
$M_{\mathrm{median}}$-$n_{\mathrm{c}}$ relation typically is shallower in the
low-resolution models. This can be seen when comparing
Fig.~\ref{fig:all-med-rho} and Fig.~\ref{fig:all-slope}a, where we used
identical turbulent driving fields.

For all four different realizations of the turbulent driving field shown in
Fig.~\ref{fig:all-slope}a-d we see a clear trend of decreasing
median mass $M_{\mathrm{median}}$ with increasing $n_{\mathrm{c}}$. We
conclude that, qualitatively, the $M_{\mathrm{median}}$-$n_{\mathrm{c}}$
relation is independent of the details of the turbulent driving field but, quantitatively, there are
significant variations.  
This is not surprising given the stochastic nature of
turbulent flows. A further discussion on this issue can be found in \citet{KLE00b} and \citet{HEI01}. 
\subsection{Dependence on the scale of turbulent driving}
Figure~\ref{fig:all-slope}e shows the results for models where the driving
scale has been changed to a lower value ($k=7..8$).  The overall
  dependency of $M_{\mathrm{median}}$ on $n_{\mathrm{c}}$ is very similar to
  the cases of large-scale turbulence. However, we note considerably larger
  uncertainties in the exact value of the slope. This holds especially for the
  phases where $30\%$ and $50\%$ of the total gas mass has been converted into
  stars. One of the reasons is
lower statistics, i.e., for the smallest critical density only one protostellar
object forms. Moreover, it has already been noted by \citet{LI03} that driving on small wavelength
results in less fragmentation. The forming small-scale density
structure is not so strongly filamentary, compared to the case of small-scale
driven turbulence. Local differences have a larger influence on
the results for driving on small wavelengths. 
Nevertheless, for most of the models the mean mass decreases with increasing
critical density.
Observational evidence suggests that real molecular clouds are driven from
large scales \citep[e.g.][]{OSS02, BRU04a}.

\section{Summary}
\label{sec:sum}
Using SPH simulations we investigate the influence of a piecewise polytropic
EOS on fragmentation of
molecular clouds. We study the case where the polytropic index $\gamma$
changes from a value below unity to one above at a critical density
$n_{\mathrm{c}}$. We consider a broad range of values of $n_{\mathrm{c}}$
 around the realistic value to determine the dependence of the mass spectrum
 on $n_{\mathrm{c}}$.

Observational
evidence predicts that dense prestellar cloud cores show rough
balance between gravity and thermal pressure \citep{BEN89, MYE91}.
Thus, the thermodynamical properties of the gas play an important role
in determining how dense star-forming regions in molecular clouds
collapse and fragment.    
Observational and theoretical studies of the thermal properties of
collapsing clouds both indicate that at densities below
$10^{-18}\,\mathrm{g\,cm^{-3}}$, roughly corresponding to
a number density of $n_{\mathrm{c}}=2.5\times10^5\,\mathrm{cm^{-3}}$, the temperature
decreases with increasing density. This is due to the strong dependence of
molecular cooling rates on density \citep{KOY00}. Therefore, the polytropic
exponent $\gamma$ is below unity in this density regime. At densities above
$10^{-18}\,\mathrm{g\,cm^{-3}}$, the gas becomes thermally coupled to
the dust grains, which then control the temperature by far-infrared
thermal emission. The balance between compressional heating and
thermal cooling by dust causes the temperature to increase again slowly with
increasing density. Thus the temperature-density relation can be
approximated with $\gamma$ above unity \citep{LAR85} in this regime. Changing
$\gamma$ from a value below unity to a value above unity results in a
minimum temperature at the critical density. \citet{LI03} showed that
gas fragments efficiently for $\gamma < 1.0$ and less efficiently for
higher $\gamma$. Thus, the Jeans mass at the critical density defines
a characteristic mass for fragmentation, which may be related to the
peak of the IMF.
 
We investigate this relation numerically by changing $\gamma$ from
$0.7$ to $1.1$ at different critical densities
$n_{\mathrm{c}}$ varying from $4.3\times10^4\,\mathrm{cm^{-3}}$ to
$4.3\times10^7\,\mathrm{cm^{-3}}$.  
A simple scaling argument based on the Jeans mass $M_\mathrm{J}$ at the
critical density $n_\mathrm{c}$ leads to $M_{\mathrm{ch}}\propto
n_{\mathrm{c}}^{-0.95}$. If there is a close relation between the average
Jeans mass and the characteristic mass of a fragment, a similar relation
should hold for the expected peak of the mass spectrum.
 Our simulations qualitatively support this hypothesis,
however, with the weaker density dependency $M_{\mathrm{ch}} \propto
n_{\mathrm{c}}^{-0.5\pm0.1}$. 
The density at which $\gamma$ changes from below unity to
above unity selects a characteristic mass scale. Consequently, the
peak of the resulting mass spectrum decreases with increasing
critical density. This spectrum not only shows a pronounced peak but also a
powerlaw tail towards higher masses. Its behavior is thus similar to the
observed IMF. 

Altogether, supersonic turbulence in self-gravitating molecular gas generates a
complex network of interacting filaments. The overall density
distribution is highly inhomogeneous. Turbulent compression sweeps up
gas in some parts of the cloud, but other regions become rarefied.
The fragmentation behavior of the cloud and its ability to form stars
depend strongly on the EOS.  
If collapse sets in, the final mass of a fragment depends not only on the
local Jeans criterion, but also on additional processes. For
example, protostars grow in mass by accretion from their surrounding material.
In turbulent clouds the properties of the gas reservoir are
continuously changing. In a dense cluster environment, furthermore,
protostars may interact with each other, leading to ejection or mass
exchange. These dynamical factors modify the resulting mass
spectrum, and may explain why the characteristic stellar mass
depends on the EOS more weakly than expected.

We also studied the effects of different
turbulent driving fields and of a smaller driving scale.  For different realizations of statistically
identical large-scale turbulent velocity fields we consistently find the
characteristic mass is decreasing with increasing critical mass. 
However, there
are considerable variations.  The influence of the natural stochastic
fluctuations in the turbulent flow on the resulting median mass is
almost as pronounced as the changes of the thermal properties of the
gas. Also when inserting turbulent energy at small wavelengths we see
the peak of the mass spectrum decreasing with increasing critical
density. 

Our investigation supports the idea that the distribution
of stellar masses depends, at least in part, on the thermodynamic state of the
star-forming gas.  If there is a low-density regime in molecular
clouds where temperature $T$ sinks with increasing density $\rho$, 
followed by a higher-density phase where $T$ increases with $\rho$, fragmentation seems likely to be favored at the transition density
where the temperature reaches a minimum. This defines a characteristic
mass scale. The thermodynamic state of interstellar gas is a result of
the balance between heating and cooling processes, which in turn are
determined by fundamental atomic and molecular physics and by chemical abundances. The derivation
of a characteristic stellar mass can thus be based on quantities and constants
that depend solely on the chemical abundances in a molecular cloud. 
The current study using a piecewise polytropic EOS
can only serve as a first step. Future work will need to consider a
realistic chemical network and radiation transfer processes in gas of varying abundances.

\acknowledgements{ We acknowledge interesting and stimulating discussions with S.~Glover and M.~Spaans. A.~K.~J. thanks the AMNH for its warm hospitality and the
  Kade Foundation for support of her visits there. A.~K.~J. and R.~S.~K. acknowledge support
  from the Emmy Noether Program of the Deutsche Forschungsgemeinschaft
  (grant no.\ KL1358/1). Y.~L. and M.-M.~M.~L. acknowledge partial support by
  NASA grants NAG5-13028 and NAG5-10103, and by NSF grant AST03-07793.}

\appendix
\section{Implementation of Sink Particles}
The parallel version of GADGET distributes the SPH particles onto the
individual processors, using a spatial domain decomposition. Thus, each
processor hosts a rectangular piece of computational volume.
If the position of a sink particle is near the boundary of this volume, the
accretion radius overlaps with domains on other processors. We therefore
communicate the data of the sink to all processors. Every processor searches
for gas particles within the accretion radius of the sink. Three criteria
determine whether the particle gets accreted or not. First, the particle
must be bound to the sink particle, i.e., the kinetic energy must be less than
the magnitude of the gravitational energy.
Second, the specific angular momentum of the particle must be less than what
is required to move on a circular orbit with radius $r_{\mathrm{acc}}$ around
the sink particle. Finally, the particle must be more tightly bound to the
candidate sink particle than to other sink particles.
Once the
central region of a collapsing gas clump exceeds a density
contrast $\Delta\rho/\rho\sim 5000$, we introduce a new sink particle. The procedure for
dynamically creating a sink particle is as follows. We search all processors
for the gas particle with the highest density. When this density is above the
threshold and when its smoothing length is less than half the accretion
radius, then the gas particle is considered to become a sink particle. If the
accretion radius around the candidate particle overlaps with another domain,
its position is sent to the other processors. Every processor searches for the
particles that exist in its domain and, simultaneously, within the accretion radius of the candidate particle. These particles and the candidate particle undergo a series of tests to
decide if they should form a sink particle. 
First, the new sink particle must be the only one within two accretion
radii. Second, the ratio of thermal energy to the magnitude of the
gravitational energy must be less than 0.5.
Third, we require that the total energy is less than zero. Finally, the
divergence of the accelerations on the particles must be less than zero.
If all these tests are passed, the particle with the highest density turns
into a sink particle with position, velocity and acceleration derived from the
center of mass values of the original gas particles. If these original particles are
distributed over several processors the center of mass values have to be communicated
correctly to the processor that hosts the new sink particle. 

Ideally, the creation of sink particles in an SPH simulation should not affect
the evolution of the gas outside its accretion radius. In practice there is
the discontinuity in the SPH particle distribution due to the hole produced by
the sink particles. This affects the pressure and viscous forces on particles
outside. We have implemented adequate boundary conditions at the "surface" of
the sink particles as described in detail in \citet{BAT95} to correct for
these effects. 

Following \citet{BAT95} we use the \citet{BOS79} standard isothermal
test case for the collapse and fragmentation of an interstellar cloud
core to check our implementation. Initially, the cloud core is
spherically symmetric with a small $m=2$ perturbation and uniformly
rotating. As gravitational collapse proceeds a rotationally supported
high-density bar builds up in the center embedded in a disk-like
structure. The two ends of the bar become gravitationally unstable,
resulting in the formation of a binary system. We see no further
subfragmentation \citep[see also,][]{TRU97}. These tests
show that the precise creation time and the mass of the sink particle
at the time of its formation can vary somewhat with the number of used
processors. We also find that simulations with different processor
numbers show small deviations in the exact positions and velocities of
the gas particles. These variations are due to the differences in the extent of
the domain on each processor. When the force on a particular particle is
computed, the force exerted by distant groups is approximated by their lowest
multipole moments. Since each processor constructs its own Barnes and Hut tree
 differences in the tree walk result in differences in the computed force. Hence, the formation mass and time of sink
particles depend on the computational setup. Nevertheless, these
differences are only at the $0.1\%$ level and the total number of
collapsing objects is not influenced by a change in the number of
processors.
\nocite{ARC91}
\nocite{SCA98b}
\bibliographystyle{aa}
\bibliography{myref}

\end{document}